\documentclass[runningheads]{llncs}
\usepackage{graphicx}
\usepackage{graphicx}
\usepackage{hyperref}
\usepackage{ifthen}
\usepackage{textcomp}
\usepackage{xcolor}
\usepackage{svg}
\usepackage{cite}
\usepackage{amsmath,amssymb,amsfonts}
\usepackage{algorithmic}
\usepackage{graphicx}
\usepackage{caption}
\usepackage{subcaption}
\usepackage{fancyvrb}
\usepackage{xcolor}

\usepackage[T1]{fontenc}
\newboolean{showcomments}
\setboolean{showcomments}{true} 
\ifthenelse{\boolean{showcomments}}
  {\newcommand{\nb}[2]{
    \fcolorbox{gray}{yellow}{\bfseries\sffamily\scriptsize#1}
    {\sf\small$\blacktriangleright$\textit{#2}$\blacktriangleleft$}
   }
   
  }
  {\newcommand{\nb}[2]{}
   
  }

\definecolor{codered}{RGB}{177,84,105}
\definecolor{codeblue}{RGB}{86,53,255}
\definecolor{codegreen}{RGB}{63,127,95}
\definecolor{codesalmon}{RGB}{250,128,114}

\begin{document}
\title{Low-Modeling of Software Systems}
\titlerunning{Low-Modeling of Software Systems}
%
\author{Jordi Cabot\inst{1,2}}

\authorrunning{J. Cabot}
%
\institute{Luxembourg Institute of Science and Technology, Esch-sur-Alzette, Luxembourg \\
\email{jordi.cabot@list.lu} \and
University of Luxembourg, Esch-sur-Alzette, Luxembourg }
\maketitle              
\begin{abstract}

There is a growing need for better development methods and tools to keep up with the increasing complexity of new software systems. New types of user interfaces, the need for intelligent components,  sustainability concerns, ... bring new challenges that we need to handle. In the last years, model-driven engineering has been key to improving the quality and productivity of software development, but models themselves are becoming increasingly complex to specify and manage. In this paper, we present the concept of \textit{low-modeling} as a solution to enhance current model-driven engineering techniques and get them ready for this new generation of software systems.  

\end{abstract}

\section{Introduction}

Current software development projects face a growing demand for advanced features. Including support for new types of user interfaces (augmented reality, virtual reality, chat and voice interfaces,...), intelligent behaviour to be able to classify/predict/recommend information based on user's input or the need to face new security and sustainability concerns, among many other new types of requirements.   

To tame this complexity, software engineers typically choose to work at a higher abstraction level~\cite{Booch18} where technical details can be ignored, at least during the initial development phases.
{\em Low-code platforms} are the latest incarnation of this trend, promising to accelerate software delivery by dramatically reducing the amount of hand-coding required. Low-code can be seen as a continuation or specific style of other model-based\footnote{Here we are referring to \emph{software models} (\emph{e.g.}, state machine diagrams), not Machine Learning models.} approaches~\cite{Cabot20}, where high-level software specifications are used to (semi)automatically generate the running software system. 

Nevertheless, even software models are becoming more and more complex due to the increasing complexity of the underlying systems being modeled. Beyond ``classical'' data and behavioural aspects we now need to come up with new models to define the new types of UIs or all the smart features of the software. Note that AI elements are hard to specify~\cite{rahimi2019toward}, architect, test and verify~\cite{RiccioJSHWT20} and low-code systems have so far paid little attention to the modeling and development of smart systems. 

In this sense, we argue for the need of \textit{low-modeling} techniques that accelerate the modeling process from which then the system will be generated.

The next sections define the concept of low-modeling (Section \ref{sec:definition}) and describe a number of low-modeling strategies (Section \ref{sec:strategies}). We then illustrate the benefits of low-modeling through a specific case study (Section \ref{sec:usecsae}) and provide a very brief introduction of an ongoing low-modeling platform project (Section \ref{sec:besser}).

\section{Low modeling definition }
\label{sec:definition}

The Forrester Report\cite{richardson2014new} states that low-code application platforms accelerate app delivery by dramatically reducing the amount of hand-coding required. Similarly, we define \emph{low-modeling} as the \textit{set of strategies that accelerate the modeling of a software system by dramatically reducing the amount of hand-modeling required}. 

Often, a low-modeling platform will also follow a low-code approach to generate the final software code from the (semi)automatically generated models. 

Low-modeling can also improve the adoption of modeling in companies and organizations. It is well accepted that the adoption challenge is a complex sociotechnical problem \cite{HutchinsonWR14}. And it is aggravated when considering that software is now being developed by multidisciplinary teams, e.g to deal with the specification and development of the intelligent components embedded in most new systems. In this sense, the goal of low-modeling is not only to increase the productivity of developer teams, but also to contribute to the democratization of software development by enabling all types of professionals to participate in the development process and even build their own applications beyond what low-code, no-code and template-based approaches offer. 

\section{Low modeling strategies }
\label{sec:strategies}

This section gives a short overview of several strategies that could be employed to put in place a low-modeling strategy. Similarly to low-code approaches where code is semi-automatically generated from ``earlier'' sources (i.e. models in that case), in a low-modeling approach we will see how models are generated also from other input sources, such as existing knowledge or (un)structured documents. 

Note that many of the techniques employed to do so are not new but they will need to be adapted and extended to cover the new types of models required to specify today's systems (e.g. its smart capabilities or new types of interfaces as exemplified in the next section). 

The list does not aim to be exhaustive. And any modeling strategy will probably employ a combination of techniques, where the right combination and specific type of low-modeling approach will depend on the specific needs of the system-to-be and the resources already available in the organization (e.g. existing information, how standard is the domain we are modeling, how simple/complex are their features, ...).

According to the low-modeling definition, the goal of these techniques is to create initial versions of models (or more complete versions of existing ones) to be then validated and refined by modeling experts. The goal is NOT to replace the need for modeling but to let modelers focus on the more creative and key aspects of the modeling activity instead of wasting time on boilerplate modeling. 

\subsection{Heuristic-based model generation}

Convention over configuration is a software design paradigm used by software frameworks that attempts to decrease the number of decisions that a developer using the framework is required to make without necessarily losing flexibility and don't repeat yourself (DRY) principles \footnote{\url{https://en.wikipedia.org/wiki/Convention_over_configuration}}. We could use this principle at the modeling level to reduce the number of modeling decisions a designer has to make (hence, being more low-modeling). 

In particular, convention over configuration and the use of heuristics can help us to create basic models for parts of the system from other existing models. A good example is the automatic generation of behavioural models \cite{albert2011generating} or user interface models \cite{rodriguez2018autocrud} from static models. The key idea is that any data model will require a number of basic CRUD (create/read/update/delete) operations to visualize and manipulate the data specified in the model. This operations can be deduced from an analysis of the static model elements and relationships by systematically applying a number of heuristics. This is similar to the scaffolding features offered by most web programming frameworks. Obviously, this generated behaviuor must then be refined and new behaviours, that go beyond the CRUD core elements, added to complete the model. Still, this represents a minor percentage of the total model size and creation time (the Pareto rule also applies here).

\begin{figure}
    \includegraphics[width=\textwidth]{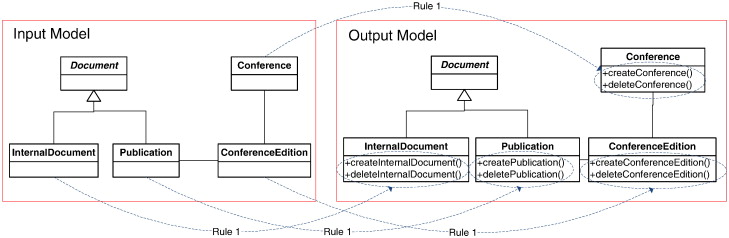}
    \caption{CRUD-driven operation generation, taken from \cite{albert2011generating}.} \label{fig:crud}
\end{figure}

\subsection{Knowledge-based model enrichment}

For many domains, there is plenty of structured knowledge already available. From simple thesaurus to general ontologies like Cyc \cite{lenat1995cyc}. This knowledge can be used to enrich a partial model with alternative concepts related to those already present in the model (e.g. based on the distance, in the ontology hierarchy, between the existing models and potentially new ones). For some domains, more specific ontologies, targeting the knowledge of that particular domain, could exist and produce better results. An even more extreme approach can involve the derivation of the target model by pruning all the superfluous concepts (for the system at hand) from an initial ontology \cite{conesa2006method}.

Knowledge can also come from previous models created as part of previous modeling projects in the same domain. Either by the same company or by others but contributed to a common model repository \cite{RoccoRIP15,france2006repository}. As before, these previous models could be compared with the current one to suggest ways to enrich it. 

Regardless of the specific method, the key idea is to reuse existing knowledge, already formalized by other individuals or whole communities, to speed up the creation of new models for the same domain. And not only that, this knowledge-reuse can also improve the overall quality. Differences between the model and the existing knowledge-bases could suggest errors in the model. These potential errors would then need to be revised by the expert so conclude whether the error is true or it is just that for this specific system we are deviating from more common specifications. 

\subsection{ML-based model inference}

The last group of techniques deal with the variety of ML techniques and applications that could help to infer models \cite{BurguenoCWZ22} from unstructured sources. This ranges from the automatic derivation of models from the textual analysis of documents to the creation of modeling assistants (similar to what GitHub copilot offers to programmers) thanks to the use of Generative AI techniques \cite{CamaraTBV23}.

As all the other domains where AI is applied, these techniques end up being the most powerful ones (as they can extract models from completely unstructured sources and with the least human intervention) but at the same time the ones that pose the highest risk as there are no guarantees in the quality of the result. They may be the fastest way to get some results but they are also the most time-consuming during the review phase. 

Note that the quality of the results largely depends on the datasets used during the ML training. It is then worth mentioning initiatives targeting the creation and curation of proper model datasets for machine learning, such as \cite{LopezIC22}.

\section{Case Study : low modeling of conversational interfaces}
\label{sec:usecsae}

As an example of how these low-modeling strategies could be combined to accelerate the development of smart components and systems, we illustrate their role in a specific scenario: the automatic development of chatbots to talk to (open) data sources.

In real-world applications, the most common data type is tabular data. With the rise of digital technologies and the exponential growth of data, the number of tabular data sources is constantly increasing and is expected to continue to do so in the future. In particular, tabular data is also the underlying mechanism used by all kinds of public administrations to publish public data sets, known as open data. Indeed, a quick search in any of the public administration open data and transparency portals reveals the large number of data sources published\footnote{Just the EU portal \url{https://data.europa.eu/} registers over 1.5M} and the popularity of CSV and other similar tabular data formats to publish those. 
Despite its importance, there is a lack of methods and tools that facilitate the exploration of tabular data by non-technical users. 

Conversational User Interfaces (CUI), embedded in chatbots and voicebots, have been proven useful in many contexts to automate tasks and improve the user experience, such as automated customer services, education and e-commerce. We believe they could also play a major role in the exploitation of tabular data sources. Until now, such chatbots for tabular data were either manually created (an option that it is not scalable) or completely relying on pure English-to-SQL translational approaches (with limited coverage and with a risk of generating wrong answers). We have been working on a new, fully automated, approach where bots are automatically derived based on an analysis of the tabular data description and contents.

The low-level technical details of the solution are described in \cite{gomez2023automatic} but in this section we expand on the key role of low-modeling in the generation of such bots. 

\subsection{From tabular data to data models}
\label{sec:preliminary}

We aim to generate CUIs to interrogate tabular data sources. Tabular data is structured into rows, each of which contains information about some observation. Each row contains the same number of cells (they could be empty), which provide values of properties of the observation described by the row. In tabular data, cells within the same column provide values for the same property of the observations described by each row \cite{w3c}.

\begin{figure}
    \includegraphics[width=\textwidth]{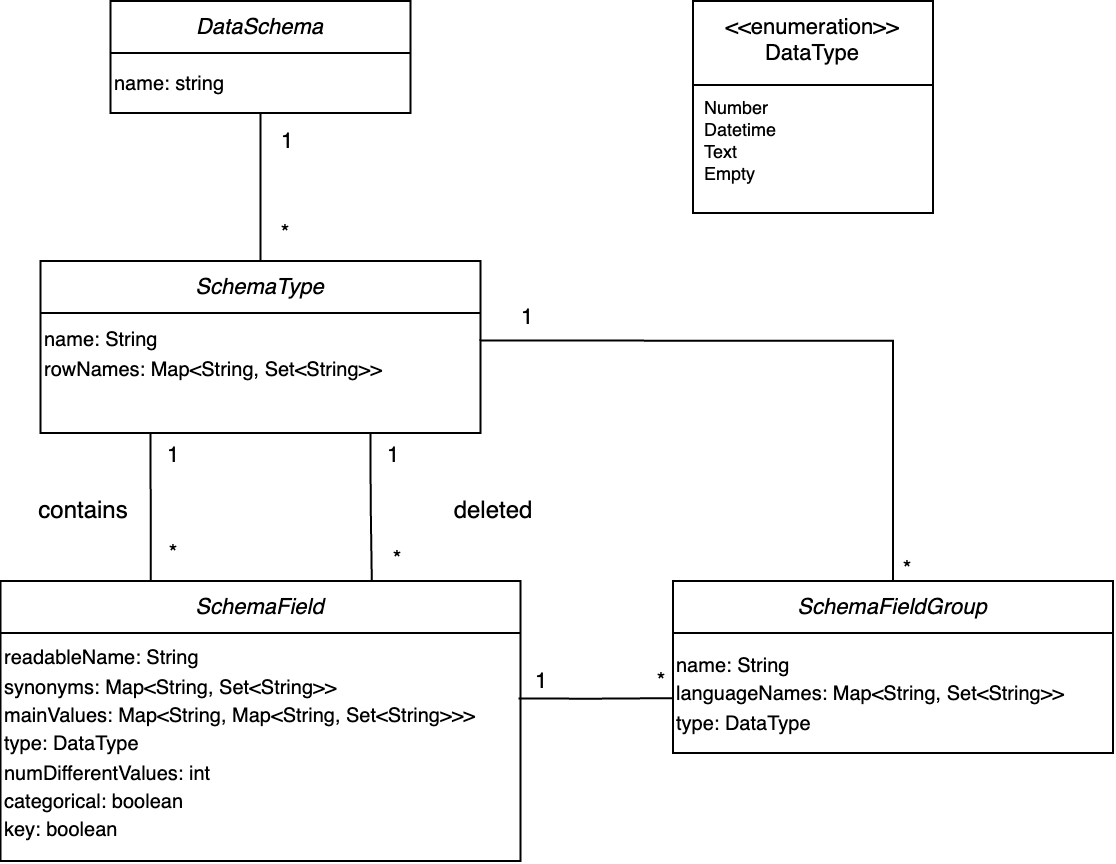}
    \caption{DataSchema metamodel.} \label{fig:ds-metamodel}
\end{figure}

A static analysis of the tabular data columns and contents, enables us to infer enough information to fill a simple data model as the one in Figure \ref{fig:ds-metamodel}. From the structure of the dataset we will gather the list of columns/fields (with their names). From the analysis of the dataset content, we will infer the data type of each field (numeric, textual, date-time,...) and its diversity (number of different values present in that specific column). Based on a predefined (but configurable) diversity threshold, we automatically classify as \textit{categorical} those fields under the threshold. 

We can enrich this data model with additional information, e.g. adding synonyms based on thesaurus to improve the bot comprehension capabilities and the use of ontologies to detect semantic relationships between the fields. 

\subsection{From data models to conversational models}

Chatbots conversation capabilities are designed as a set of \emph{intents}, where each intent represents a possible user goal when interacting with the bot. The bot then tries to match any user utterance (i.e., user's input question) to one of its intents. As part of the match, one or more parameters (also called \textit{ entities} in bot terminology) in the utterance can also be recognized, in a process known as \textit{named entity recognition} (NER). 
When there is a match, the bot back-end executes the required behaviour.

\begin{figure}[htb!]
	\centering
	\includegraphics[width=290pt]{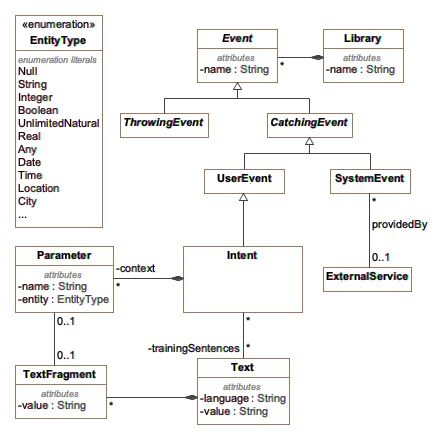}
	\caption{Intent package metamodel from \cite{planas2021towards}}
	\label{fig:IntentPackage}
\end{figure}

We have then defined a set of heuristics\cite{gomez2023automatic} that are iteratively applied to the data model to generate a conversation model compliant with the conversation metamodel partially depicted in Figure \ref{fig:IntentPackage}.

\subsection{From conversational models to the actual chatbot}

This last step is straightforward as it just involves a model-to-text transformation to go from an initial tabular data source to an actual running chatbot via a couple of intermediate models in a fully automated way thanks to the use of a combination of low-modeling strategies.

Note that even if an initial working version of the models is automatically generated, all models are explicit and can be manually refined at every step if needed.

\section{BESSER: a low-code low-modeling platform}
\label{sec:besser}

We will continue exploring these ideas as part of the \textit{BESSER} platform, a low-code and low-modeling platform to speed up the definition of high quality smart software. BESSER is a 5-years project. As part of the project results, we are developing an open-source platform implementing the project results, available in our GitHub organization \footnote{\url{https://github.com/besser-pearl}}.

\begin{figure*}
    \centering
    \includegraphics[width=\textwidth]{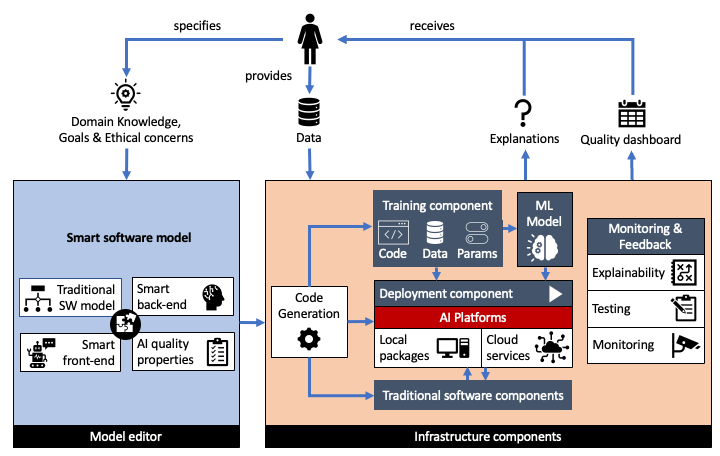}
    \caption{The low-code architecture proposed in \cite{cabot2022low}}
    \label{fig:arch}
\end{figure*}

BESSER will extend the low-code architecture depicted in Figure \ref{fig:arch} with low-modeling components able to generate (partial) versions of all the models for smart software from (un)structured data sources using a variety of static analysis, knowledge engineering and ML-based model inference techniques such as those mentioned in the previous sections. 

Note how these inference techniques will need to target each type of model separately (the traditional models, the smart front-end ones such as the conversational models we have just seen, the smart back-end ones, etc) but also the interaction between them to ensure they behave in a consistent way.

\section{Conclusions and further work}
\label{section:conclusions-and-future-work}
This paper has introduced the concept of low-modeling and given examples of different types of low-modeling strategies that could be used to improve the development of complex systems, such as smart software systems, following a model-driven approach. This is just a first step in this direction, as the community is still proposing new types of languages and models to cover all aspects required to precisely define such smart systems. New low-modeling strategies will be needed to accelerate the development of these new kinds of models.

Beyond this challenge, we also believe that modeling languages themselves will need to be more flexible and integrate as first-class elements, new modeling concerns. For instance, we believe uncertainty modeling \cite{TroyaMBV21} should be considered a first-level concern. Not only AI systems are full of uncertainty per se (at all levels, at the data level, at the ML model level,...) but the result of most low-modeling strategies comes with its own level of confidence. Moreover, low-modeling strategies may generate partial models that may need to dynamically evolve if the environment changes. Finally, low-modeling will need to go beyond the inference of the key data and behavioural elements of the system being modeled and become able to also suggest initial models for other system aspects such as the security concerns or even the ethical constraints.

\subsubsection*{Acknowledgements.}
This project is supported by the Luxembourg National Research Fund (FNR) PEARL program, grant agreement 16544475.

\bibliographystyle{splncs04}
\bibliography{refs}

\begin{thebibliography}{10}
\providecommand{\url}[1]{\texttt{#1}}
\providecommand{\urlprefix}{URL }
\providecommand{\doi}[1]{https://doi.org/#1}

\bibitem{albert2011generating}
Albert, M., Cabot, J., G{\'o}mez, C., Pelechano, V.: Generating operation
  specifications from uml class diagrams: A model transformation approach. Data
  \& Knowledge Engineering  \textbf{70}(4),  365--389 (2011)

\bibitem{Booch18}
Booch, G.: The history of software engineering. {IEEE} Softw.  \textbf{35}(5),
  108--114 (2018). \doi{10.1109/MS.2018.3571234}

\bibitem{BurguenoCWZ22}
Burgue{\~{n}}o, L., Cabot, J., Wimmer, M., Zschaler, S.: Guest editorial to the
  theme section on ai-enhanced model-driven engineering. Softw. Syst. Model.
  \textbf{21}(3),  963--965 (2022). \doi{10.1007/s10270-022-00988-0},
  \url{https://doi.org/10.1007/s10270-022-00988-0}

\bibitem{Cabot20}
Cabot, J.: Positioning of the low-code movement within the field of
  model-driven engineering. In: Companion Proc. of {MODELS}'20. pp. 76:1--76:3.
  {ACM} (2020). \doi{10.1145/3417990.3420210}

\bibitem{cabot2022low}
Cabot, J., Claris{\'o}, R.: Low code for smart software development. IEEE
  Software  \textbf{40}(1),  89--93 (2022)

\bibitem{CamaraTBV23}
C{\'{a}}mara, J., Troya, J., Burgue{\~{n}}o, L., Vallecillo, A.: On the
  assessment of generative {AI} in modeling tasks: an experience report with
  chatgpt and {UML}. Softw. Syst. Model.  \textbf{22}(3),  781--793 (2023).
  \doi{10.1007/s10270-023-01105-5},
  \url{https://doi.org/10.1007/s10270-023-01105-5}

\bibitem{conesa2006method}
Conesa, J., Olive, A.: A method for pruning ontologies in the development of
  conceptual schemas of information systems. In: Journal on Data Semantics V,
  pp. 64--90. Springer (2006)

\bibitem{france2006repository}
France, R., Bieman, J., Cheng, B.H.: Repository for model driven development
  (remodd). In: International Conference on Model Driven Engineering Languages
  and Systems. pp. 311--317. Springer (2006)

\bibitem{gomez2023automatic}
Gomez, M., Cabot, J., Clarisó, R.: Towards the automatic generation of
  conversational interfaces to facilitate the exploration of tabular data
  (2023)

\bibitem{HutchinsonWR14}
Hutchinson, J.E., Whittle, J., Rouncefield, M.: Model-driven engineering
  practices in industry: Social, organizational and managerial factors that
  lead to success or failure. Sci. Comput. Program.  \textbf{89},  144--161
  (2014). \doi{10.1016/j.scico.2013.03.017},
  \url{https://doi.org/10.1016/j.scico.2013.03.017}

\bibitem{lenat1995cyc}
Lenat, D.B.: Cyc: A large-scale investment in knowledge infrastructure.
  Communications of the ACM  \textbf{38}(11),  33--38 (1995)

\bibitem{LopezIC22}
L{\'{o}}pez, J.A.H., Izquierdo, J.L.C., Cuadrado, J.S.: Modelset: a dataset for
  machine learning in model-driven engineering. Softw. Syst. Model.
  \textbf{21}(3),  967--986 (2022). \doi{10.1007/s10270-021-00929-3},
  \url{https://doi.org/10.1007/s10270-021-00929-3}

\bibitem{planas2021towards}
Planas, E., Daniel, G., Brambilla, M., Cabot, J.: Towards a model-driven
  approach for multiexperience ai-based user interfaces. Software and Systems
  Modeling  \textbf{20},  997--1009 (2021)

\bibitem{rahimi2019toward}
Rahimi, M., Guo, J.L., Kokaly, S., Chechik, M.: Toward requirements
  specification for machine-learned components. In: 2019 IEEE 27th
  International Requirements Engineering Conference Workshops (REW). pp.
  241--244 (2019). \doi{10.1109/REW.2019.00049}

\bibitem{RiccioJSHWT20}
Riccio, V., Jahangirova, G., Stocco, A., Humbatova, N., Weiss, M., Tonella, P.:
  Testing machine learning based systems: a systematic mapping. Empir. Softw.
  Eng.  \textbf{25}(6),  5193--5254 (2020). \doi{10.1007/s10664-020-09881-0},
  \url{https://doi.org/10.1007/s10664-020-09881-0}

\bibitem{richardson2014new}
Richardson, C., Rymer, J.R., Mines, C., Cullen, A., Whittaker, D.: New
  development platforms emerge for customer-facing applications. Forrester:
  Cambridge, MA, USA  \textbf{15} (2014)

\bibitem{RoccoRIP15}
Rocco, J.D., Ruscio, D.D., Iovino, L., Pierantonio, A.: Collaborative
  repositories in model-driven engineering. {IEEE} Softw.  \textbf{32}(3),
  28--34 (2015). \doi{10.1109/MS.2015.61},
  \url{https://doi.org/10.1109/MS.2015.61}

\bibitem{rodriguez2018autocrud}
Rodriguez-Echeverria, R., Preciado, J.C., Sierra, J., Conejero, J.M.,
  Sanchez-Figueroa, F.: Autocrud: Automatic generation of crud specifications
  in interaction flow modelling language. Science of Computer Programming
  \textbf{168},  165--168 (2018)

\bibitem{TroyaMBV21}
Troya, J., Moreno, N., Bertoa, M.F., Vallecillo, A.: Uncertainty representation
  in software models: a survey. Softw. Syst. Model.  \textbf{20}(4),
  1183--1213 (2021). \doi{10.1007/s10270-020-00842-1},
  \url{https://doi.org/10.1007/s10270-020-00842-1}

\bibitem{w3c}
W3C: Model for tabular data and metadata on the web  (2015)

\end{thebibliography}

\end{document}